\documentclass[aps]{revtex4}

\usepackage[utf8]{inputenc} % allow utf-8 input
\usepackage[T1]{fontenc}    % use 8-bit T1 fonts
\usepackage{hyperref}       % hyperlinks
\usepackage{url}            % simple URL typesetting
\usepackage{booktabs}       % professional-quality tables
\usepackage{amsfonts}       % blackboard math symbols
\usepackage{nicefrac}       % compact symbols for 1/2, etc.
\usepackage{microtype}      % microtypography
\usepackage{lipsum}
\usepackage{verbatim}

\usepackage{amsmath}
\usepackage{graphicx,epstopdf,subfigure,color,hyperref}
\usepackage{mathtools}
\usepackage{mathrsfs}
\usepackage{amsthm}
\usepackage{bbold}
\usepackage{bm}
\usepackage{float}
\usepackage{textcomp}
\usepackage{setspace}
\usepackage{psfrag}
\usepackage[normalem]{ulem}
\usepackage{epsf}
\usepackage{lmodern}

\newcommand{\defeq}{\vcentcolon=}

\providecommand{\definitionname}{Definition}

\theoremstyle{definition}
\newtheorem{defn}{\protect\definitionname}

\begin{document}
\title{Double Slit with an Einstein--Podolsky--Rosen Pair}

\author{Bar Y. Peled}
\affiliation{Center for Quantum Information Science and Technology \& Faculty
	of Engineering Sciences, Ben-Gurion University of the Negev, Beersheba
	8410501, Israel}
\author{Amit Te'eni}%
\affiliation{Faculty of Engineering and the Institute of Nanotechnology and Advanced
	Materials, Bar Ilan University, Ramat Gan 5290002, Israel}
\author{Danko Georgiev}
\affiliation{Institute for Advanced Study,
	30 Vasilaki Papadopulu Str., Varna 9010, Bulgaria}
\author{Eliahu Cohen}
\affiliation{Faculty of Engineering and the Institute of Nanotechnology and Advanced
	Materials, Bar Ilan University, Ramat Gan 5290002, Israel}
\author{Avishy Carmi}
\affiliation{Center for Quantum Information Science and Technology \& Faculty
	of Engineering Sciences, Ben-Gurion University of the Negev, Beersheba
	8410501, Israel}

\begin{abstract}
In this somewhat pedagogical paper we revisit complementarity relations in bipartite quantum systems. Focusing on continuous variable systems, we examine the influential class of EPR-like states through a generalization to Gaussian states and present some new quantitative relations between entanglement and local interference within symmetric and asymmetric double-double-slit scenarios. This approach is then related to ancilla-based quantum measurements, and weak measurements in particular. Finally, we tie up the notions of distinguishability, predictability, coherence and visibility while drawing some specific connections between them.
\end{abstract}

% keywords can be removed

\maketitle

\section{Introduction}
Distinguishability (or predictability) of paths in quantum interference experiments is closely linked with entanglement. For instance, to gain ``which-path'' information in a double-slit experiment, one should employ some kind of a measuring pointer (be it responsible for a strong, weak, partial measurement, etc.). The process of measuring the system's path involves entangling it with the device. For simplicity, one could assume the system is already entangled with another, similar system (in the desired basis); then, the ``which-path'' information is contained in the correlations between the two systems - i.e., knowing which slit the particle went through in one system supplies information regarding which slit the other particle went through in the other system.

This approach was utilized by Jaeger et al. in~\cite{jaeger1993complementarity}, where the authors proposed a quantitative measure for two-particle interference, denoted by $v_{12}$, and proved a duality relation between one-particle interference visibility and two-particle interference visibility: $ v_i^2 + v_{12}^2 \leq 1 $ (where $v_i$ denotes the $i$th particle's interference visibility). This and other similar duality relations were studied in ~\cite{greenberger1988simultaneous,jaeger1995two,englert1996fringe,franson1989bell}.
For general definitions and discussion of interferometric quantities in multiple-slit scenarios, please see Appendix \ref{measures}.

However, it seems to us that most of these works did not emphasize enough the state- and geometry-dependent quantitative relations between entanglement and interference of continuous variables (CV) such as position and momentum (or the corresponding quadratures of the electromagnetic field). The notions of nonlocality and entanglement in continuous variable systems are intricate and subtle. Working in infinite-dimensional Hilbert spaces necessitates new theoretical methods (for entanglement and even quantumness quantification), new practical methods (like homodyne and heterodyne detectors), new tools (like Wigner functions, Q- and P-representations) and new conceptual ideas, but they also enable new opportunities for quantum communication and computation.

Since the first entangled state to reach world-wide acclaim was the continuous variable EPR state~\cite{EPR35,schrodinger1935discussion}, it is natural to ask whether complementarity relations similar to the aforementioned ones can be formulated in EPR-like states using continuous variables. But in fact, our analysis here is more general, considering asymmetric configurations of Gaussian states, which reach the EPR state as a special, limiting case. Moreover, these CV states seem to offer richer relations between local and nonlocal observables than their discrete variables counterparts.

While discrete variable systems benefit from high-fidelity operations, they are limited by the imperfect generation and detection of single quanta (e.g. photons), as well as the absence of deterministic interactions of single quanta \cite{Masada2015}. In contrast, encoding of quantum information in continuous variables can achieve deterministic, unconditional operation of quantum protocols, albeit at the expense of lower fidelities \cite{Masada2015,Pirandola2015}.
Although the high efficiency and unconditional preparation of CV entanglement is paid for with imperfection of the entanglement \cite{Braunstein2005,Eisert2003,Serafini2017}, the prepared Gaussian entangled states approach an ideal EPR state in the limit of infinite squeezing \cite{Adesso2007}.
Thus, the practicality of continuous variables in quantum protocols is due to currently available highly efficient sources of squeezed laser light and existing techniques for preparing, unitarily manipulating, and measuring entangled quantum states, with the use of continuous quadrature amplitudes of the quantized electromagnetic field \cite{Braunstein2005}. 
It is worth noting that in this paper we are mainly focused on theoretical aspects; experimental tests of complementarity using quantum optics may be found in \cite{scully1991quantum,englert1992quantum,herzog1995complementarity,bertet2001complementarity,braig2003experimental,jacques2008illustration}, as well as a recent experiment using complementarity for nonlocal erasure of phase objects \cite{APL19}.

In Sec. \ref{partially} we define a family of EPR-like states, modeled using Gaussian wavepackets, where the strength of entanglement may be controlled by some parameter $\theta$.
We propose a quantitative measure for $v_{12}$, and %use it to study the interplay between one-particle and two-particle interference pattern visibility in continuous-variable systems.
find a relation analogous to the one-particle and two-particle interference complementarity. In the limit of zero-width Gaussian wavepackets (i.e. delta functions), our proposed system effectively reduces to a two-dimensional system, reproducing results similar to those shown in~\cite{greenberger1988simultaneous,jaeger1993complementarity}.

Furthermore, in Sec. \ref{asymmetric} we define an ``asymmetric'' state, i.e. one where Alice's system may have different parameters than Bob's, in terms of how wide the slits are and to which extent the wavepacket is localized. Again we derive an expression for the one-particle interference pattern visibility, and study its relation to the family of states we defined in Sec. \ref{partially}.
We conclude by discussing the generality of our work.

\section{Model for entangled double-slit experiment using Gaussian wave packets}

\begin{figure}[t]
	\centering
	\includegraphics[width=\textwidth]{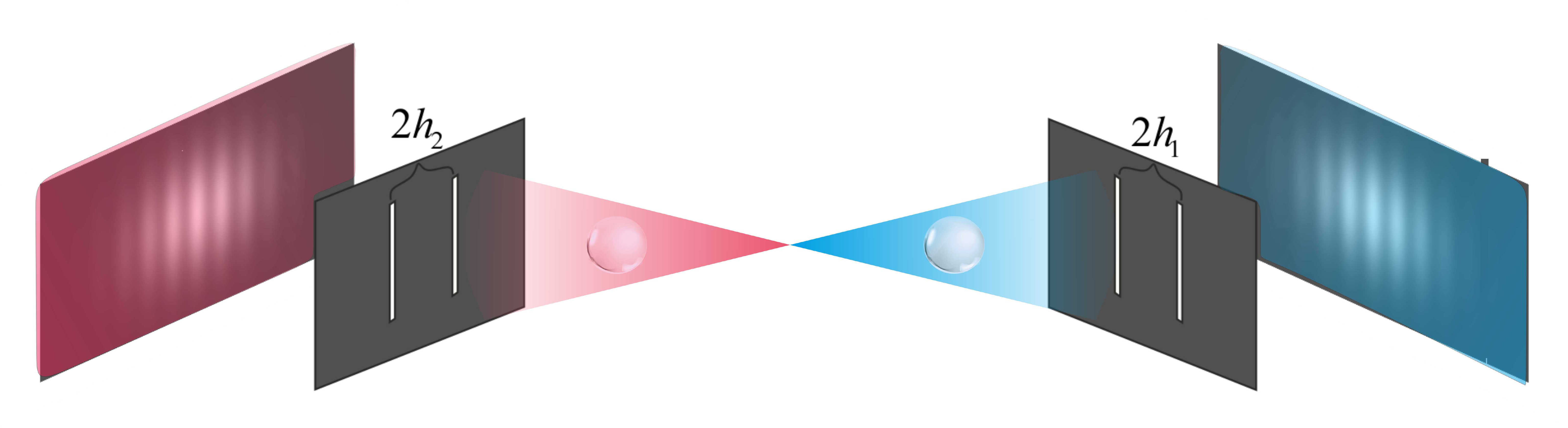}
	\caption{The proposed ``double-double-slit'' setup: each particle of an entangled two-mode Gaussian state $\psi_{\theta} \left( x_1, x_2 \right)$ \eqref{psi_theta} is sent through a double slit with a screen behind it. The amount of quantum entanglement between the two particles is controlled by a parameter $\theta \in \left[ 0, \frac{\pi}{4} \right]$. The experiment is repeated many times with the same initial state, and the positions where the particles hit the screen are registered to infer the one-particle and two-particle probability distributions. Generally, the distance between each pair of slits may vary. The plotted probability distributions on the screens are for $\theta=\frac{\pi}{4}$.}
	\label{fig:DDS}
\end{figure}

In this section we shall introduce a wavefunction which describes an \textit{entangled} double-slit experiment, based on the famous EPR state.
In the original EPR paper~\cite{EPR35}, the following state is presented
\begin{equation}\label{EPR}
\psi_{EPR} \left( x_1, x_2 \right) = \int_{-\infty}^{\infty} dx \, \delta \left( x_1 - x \right) \delta \left( x- x_2 + x_0 \right).
\end{equation}
For simplicity, we shall choose $ x_0 = 0 $. However, the above state does not reside in the Hilbert space $L^2$ of square integrable functions. Recalling that the delta function may be realized as a limit of Gaussian functions
\begin{equation}\label{delta_Gauss}
\lim_{\Delta \rightarrow 0} \frac{1}{ \left( \pi \Delta^2 \right)^{1/2} } e^{ -\left( x- x' \right)^2 / \Delta^2 } = \delta \left( x -x' \right),
\end{equation}
we substitute the delta functions in \eqref{EPR} with the LHS of \eqref{delta_Gauss}, taken with finite $\Delta$. This yields the following wavefunction
\begin{equation}
\psi_G \left( x_1, x_2 \right) = \frac{a}{\pi} \int_{-\infty}^{\infty} dx \, e^{-a \left( x_1 - x \right)^2 } e^{-a \left( x_2 - x \right)^2 },
\end{equation}
where $ a \vcentcolon= 1 / \Delta^2 $. Note that \eqref{EPR} is reproduced in the limit $ a \rightarrow \infty $.
Now, we wish to pass both particles through a double-slit, i.e. we are interested only in Gaussians centralized at one of two separate pairs of slits, located at $ x_i = \pm h_i $ for $i=1,2$ (see Fig.~\ref{fig:DDS} for an illustration). Thus, immediately after passing through the double-slit, our initial wavefunction becomes
\begin{eqnarray}
\psi \left( x_1, x_2 \right) = A \left( e^{-a\left(x_1-h_1\right)^2 } e^{-a \left( x_2-h_2 \right)^2}  + e^{ -a \left( x_1+h_1 \right)^2} e^{ -a \left( x_2+h_2  \right)^2 } \right),
\label{psi_0_entangled}
\end{eqnarray}
where $A = \sqrt{ \frac{a/\pi}{ 1 + e^{-2a \left( h_1^2 + h_2^2 \right) } } } $ is a normalization factor.

\section{A partially-entangled double-slit experiment}\label{partially}
In this section, we shall construct a model for a partially-entangled double-slit experiment. Following Sec. III of \cite{jaeger1993complementarity}, we start from \eqref{psi_0_entangled}, substitute $ h_1 = h_2 = h $, construct an ``opposite'' wavefunction and take a normalized superposition of the two, yielding
\begin{align}
& \psi_{\theta} \left( x_1, x_2 \right) = \frac{ A_{\theta} }{A}  \left[ \psi \left( x_1, x_2 \right) \cos \theta + \psi \left( x_1, -x_2 \right) \sin \theta \right],
\label{psi_theta}
\end{align}
where $\theta \in \left[ 0, \frac{\pi}{4} \right]$ is a parameter controlling the amount of entanglement (note the entanglement strength decreases as $\theta$ increases), and
\begin{equation}
\frac{1}{A_{\theta}^2} = \frac{\pi}{a} \left[ 1 + 2e^{-2ah^2} \sin \left( 2\theta \right) + e^{-4ah^2} \right].
\end{equation}
Propagation of a wave from the near to the far field transforms an initial state into its Fourier transform. Thus, the position wavefunction in some time $t$ such that the wave has propagated to reach a screen located in the far field domain, is proportional to the initial momentum wavefunction. This implies the momentum wavefunction may be used to define and compute the interferometric quantities we wish to consider (i.e. one- and two-particle visibility). The momentum wavefunction and related derivations appear in Appendix \ref{wigner_partial}.
The momentum probability density is
\begin{align}
f_{p_1, \theta} \left( p_1 \right) = \sqrt{ \frac{ 2 \pi }{ a } } \pi B_{\theta} e^{ -\frac{ p_1^2}{ 2a } } \left( \left[ 1 + e^{-2ah^2} \cos \left( 2h p_1 \right) \right] + \sin \left( 2\theta \right) \left[ e^{ -2ah^2 } + \cos \left( 2h p_1 \right) \right] \right).
\label{f_p1_theta}
\end{align}
Taking a close look at \eqref{f_p1_theta}, we notice that this probability density function is a sum of two parts: one which is independent of $p_1$ (up to normalization), and an oscillating part, with amplitude proportional to $ \sin \left( 2 \theta \right) $. This allows us to write down expressions for the ``envelopes'' of these oscillations
\begin{eqnarray}
&& env_+ \left( p_1 \right) = \pi \sqrt{ \frac{ 2 \pi }{ a } } B_{\theta} e^{ -\frac{ p_1^2}{ 2a } } \left( 1 + e^{ -2ah^2 } \right) \left[ 1 + \sin \left( 2\theta \right) \right] \\
&& env_- \left( p_1 \right) = \pi \sqrt{ \frac{ 2 \pi }{ a } } B_{\theta} e^{ -\frac{ p_1^2}{ 2a } } \left( 1 - e^{ -2ah^2 } \right) \left[ 1 - \sin \left( 2\theta \right) \right].
\end{eqnarray}
These were computed by substituting $ \pm1 $ for the cosine. However,
it should be noted that because of the Gaussian, the extreme points of $ f_{p_1, \theta} $ are not necessarily located in the extreme points of $ \cos \left( 2h p_1 \right) $. Thus, the above envelopes are merely approximations. They are ``good'' approximations if the Gaussian changes much slower than the cosine, i.e. $ ah^2 \gg 1 $.

Now we wish to construct measures for the strength of interference.
First, we compute the ratio between the envelopes
\begin{equation}
r_{\theta} \triangleq \frac{ env_+ \left( p_1 \right)}{ env_- \left( p_1 \right) } = \frac{1 + e^{ -2ah^2 } }{1 - e^{ -2ah^2 } } \cdot \frac{ 1 + \sin \left( 2\theta \right) }{ 1 - \sin \left( 2\theta \right) } = \coth \left( ah^2 \right) \frac{ 1 + \sin \left( 2\theta \right) }{ 1 - \sin \left( 2\theta \right) },
\end{equation}
where $ \coth $ denotes the hyperbolic cotangent. The envelopes also yield the well-known visibility measure
\begin{equation}
V_{\theta} = \frac{ env_+ \left( p_1 \right) - env_- \left( p_1 \right) }{ env_+ \left( p_1 \right) + env_- \left( p_1 \right) } = \frac{ e^{-2ah^2} + \sin \left( 2\theta \right) }{  1 + e^{-2ah^2} \sin \left( 2\theta \right) }.
\label{V_theta}
\end{equation}
Note that $ \sin \left( 2\theta \right) \leq V_\theta \leq 1 $, i.e. $ \sin \left( 2\theta \right) $ serves as a lower bound to the interference visibility. The more localized the Gaussian wavepackets, the closer the visibility is to $ \sin \left( 2\theta \right) $, which is also the value of the one-particle visibility in Sec. III of \cite{jaeger1993complementarity}. %This is unsurprising, since in the limit $ a \rightarrow \infty $ the Gaussian functions turn into delta functions and each particle is effectively a two-level system, thus correctly described by the analysis done there.

Next we compute the momentum joint probability distribution
\begin{align}
& f_{p_1 p_2, \theta} \left( p_1, p_2 \right) = \nonumber\\
& = B_\theta \frac{\pi}{a} e^{- \frac{p_1^2 + p_2^2 }{ 2a} } \left[ 1+ \cos^2 \theta \cos \left( 2h p_+ \right) +2 \sin \left( 2 \theta \right) \cos \left( h p_+ \right)  \cos \left( h p_- \right) + \sin^2 \theta  \cos \left( 2h p_- \right) \right],
\label{f_p1_p2}
\end{align}
where $ p_\pm \defeq p_1 \pm p_2 $.
Again, motivated by~\cite{jaeger1993complementarity}, we wish to define the two-particle visibility. To do so, we may construct the following ``corrected'' joint probability
\begin{equation*}
\bar{F}_\theta \left( p_1, p_2 \right) = f_{p_1 p_2, \theta} \left( p_1, p_2 \right) - f_{p_1, \theta} \left( p_1 \right) f_{p_2, \theta} \left( p_2 \right) + f_{p_1, \theta=0} \left( p_1 \right) f_{p_2, \theta=0} \left( p_2 \right),
\end{equation*}
where the ``correction'' term $ f_{p_1, \theta=0} \left( p_1 \right) f_{p_2, \theta=0} \left( p_2 \right) $ is simply the value of $ f_{p_1, \theta} \left( p_1 \right) f_{p_2, \theta} \left( p_2 \right) $ in the maximally-entangled case, analogous to the term $\frac{1}{4}$ added in~\cite{jaeger1993complementarity}.

\begin{figure}[t]
	\centering
	\includegraphics[width=140mm]{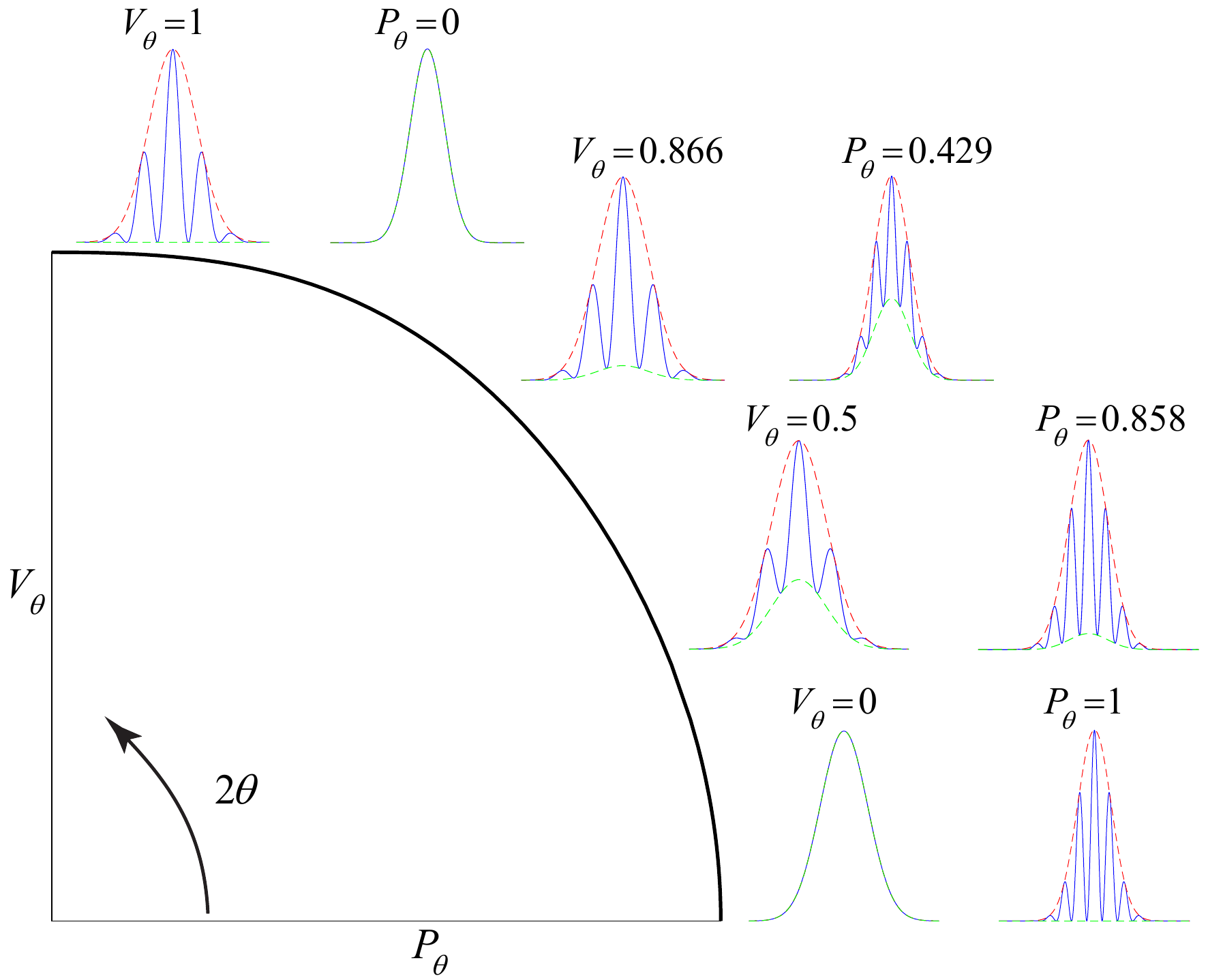}
	\caption{One-particle visibility $V_\theta$ \eqref{V_theta} and two-particle visibility (predictability) $P_\theta$ \eqref{P_theta} computed from the corresponding quantum probability distributions $f_{p_1,\theta}(p_1)$ \eqref{f_p1_theta} and $\tilde{F}_{\theta}(p_{+},p_{-}=0)$ \eqref{nicer} for different strengths of entanglement $\theta = 0, \pi/12, \pi/6 $ and $ \pi / 4$. This plot illustrates the smooth transition from zero visibility and maximal predictability, to maximal visibility and zero predictability, as the entanglement diminishes. The bold line depicts a polar curve of $ P_\theta^2 + V_\theta^2 $.}
	\label{fig:complementarity}
\end{figure}

However, this is not precisely what we do. Instead, we take the added term with the same normalization factor $B_\theta$ as the subtracted one. This yields a much ``nicer'' expression with nearly the same value
\begin{align}
\tilde{F}_\theta \left( p_{+}, p_{-} \right)
= & \frac{e^{-\frac{p_+^2 + p_-^2}{4a} } }{ C_\theta } \left[ e^{4 a h^2} + \cos^2 \left( 2\theta \right) +2 \sin^2 \left( 2 \theta \right) e^{2ah^2} \cos \left( h p_+ \right) \cos \left( h p_- \right) \right] + \nonumber\\
& + \frac{1}{2} \frac{e^{-\frac{p_+^2 + p_-^2}{4a} } }{ C_\theta } \left( 1+ \cos^2 \left( 2\theta \right) e^{4ah^2} + \cos \left( 2 \theta \right) \left[ 1+ e^{4ah^2} + 2 \sin \left( 2 \theta \right) \right] \right) \cos \left( 2h p_+ \right) + \nonumber\\
& + \frac{1}{2} \frac{e^{-\frac{p_+^2 + p_-^2}{4a} } }{ C_\theta } \left( 1+ \cos^2 \left( 2\theta \right) e^{4ah^2} - \cos \left( 2 \theta \right) \left[ 1+ e^{4ah^2} + 2 \sin \left( 2 \theta \right) \right] \right) \cos \left( 2h p_- \right),
\label{nicer}
\end{align}
where $ C_\theta \defeq 8\pi a \left[ \cosh \left( 2ah^2 \right) + \sin \left(2 \theta \right) \right]^2 $.
Since the trigonometric functions are all non-negative in our domain of interest ($ 0 \leq \theta \leq \pi /4 $), $ \cos \left( 2h p_+ \right) $ has a larger amplitude than $ \cos \left( 2h p_- \right) $. Thus, the oscillations are dominated by $p_+$, implying the envelopes are naturally defined by examining the section $p_- = 0$, i.e. $p_1 = p_2$.
The upper envelope is obtained by making the additional substitution $ hp_+ = 2\pi n, n \in \mathbb{Z} $
\begin{equation}
env_+ \left(p_+\right) = N_\theta e^{-\frac{p_+^2}{4a} } \left[ 2 + 2 e^{4ah^2} -\sin^2 \left( 2 \theta \right) \left( 1- e^{2ah^2} \right)^2 \right],
\label{key-2}
\end{equation}
and the lower envelope by $ hp_+ = \pi/2 +2\pi n, n \in \mathbb{Z} $
\begin{equation}
env_- \left(p_+\right) = N_\theta e^{-\frac{p_+^2}{4a} } \left( e^{4ah^2} -\cos \left( 2\theta \right) \left[ 1+ e^{4ah^2} +2 \sin \left( 2\theta \right) -\cos \left( 2\theta \right) \right] \right),
\label{key-3}
\end{equation}
where $ N_\theta = 1 / C_\theta = \frac{1}{8\pi a \left[ \cosh \left( 2ah^2 \right) + \sin \left(2 \theta \right) \right]^2} $. Again, the envelopes are merely approximations, becoming more precise as $ ah^2 $ grows larger. Thus we are motivated to come up with even simpler expressions for the envelopes, by applying approximations valid for $ ah^2 \gg 1 $
\begin{equation}
env_+ \left(p_+\right) \approx N_\theta e^{-\frac{p_+^2}{4a} } e^{4ah^2} \left[ 1+ \cos^2 \left( 2\theta \right) \right]
, \qquad env_- \left(p_+\right) \approx N_\theta e^{-\frac{p_+^2}{4a} } e^{4ah^2} \left[ 1 -\cos \left( 2\theta \right) \right].
\label{key-4}
\end{equation}
Thus, we obtain the following expression for the two-particle interference $ P_{\theta}$
\begin{equation}
P_{\theta} = \frac{ \cos \left( 2\theta \right) \cos^2 \theta }{ 1- \cos \left( 2\theta \right) \sin^2 \theta }.
\label{P_theta}
\end{equation}
Taking the simple expression for the visibility, $ V_\theta = \sin \left( 2 \theta \right) $, we obtain
\begin{align}
P_{\theta}^2 + V_{\theta}^2 \leq 1,
\label{P_V}
\end{align}
since $ P_{\theta} \leq \cos \left( 2\theta \right) $. Note that $ \cos \left( 2\theta \right) $ is the value of the two-particle interference visibility in Sec. III of \cite{jaeger1993complementarity}. The complementarity \eqref{P_V} is illustrated in Fig. \ref{fig:complementarity}.

The significance of inequalities of the form \eqref{P_V} is that they link a locally-measured property (one-particle visibility) and a nonlocal property (two-particle visibility). A closely related inequality is illustrated in \cite{carmi2019relativistic}
\begin{equation}
\left( \frac{\mathscr{B} }{2 \sqrt{2} } \right)^2 + |\eta|^2 \leq 1,
\label{key-6}
\end{equation}
where $ \mathscr{B} $ is the Bell-CHSH parameter -- again a measure of nonlocality; and $ \eta $ is the quantum Pearson correlation between Alice's observables, i.e. a property one may infer from measurements on Alice's system.

\section{Model for asymmetric entangled double-slit experiment}\label{asymmetric}
In this section, we consider
a generalization of \eqref{psi_0_entangled}. Recalling the aforementioned analogy between the double-double-slit and two measurement pointers measuring the entangled pair of particles, we note that Bob's system might be different from Alice's system in several aspects. Bob's pointer state could be implemented by an electron \cite{Pan2019a}, cold ion \cite{Pan2019b} or any other quantum system. Moreover, his measuring system could correspond to variable strength coupling and could even have many degrees of freedom. However, we are only interested in one continuous degree of freedom out of the above giving rise to a pair of operators denoted by $x_2, p_2$ (not necessarily position and momentum), obeying the canonical commutation relations $ \left[ x_2, p_2 \right]=i $. We therefore consider now asymmetric interactions corresponding to different measurements on the slits.
Applying these interactions results in the following state
\begin{equation}
\psi \left( x_1, x_2 \right) = M \left( e^{-a\left(x_1-h_1 \right)^2 } e^{-b \left( x_2-h_2 \right)^2}  + e^{ -a \left( x_1+h_1 \right)^2} e^{ -b \left( x_2+h_2  \right)^2 } \right), \label{asym}
\end{equation}
that is, Bob's two slits are not in the same distance as Alice's two slits, and his Gaussians have different widths than Alice's ($M$ is a normalization constant). The Wigner function of this state appears in Appendix \ref{Wigner_asym}.
We begin by writing down the wavefunction in momentum space:
\begin{eqnarray}
\psi \left(p_1, p_2 \right) = \frac{M}{2\sqrt{ab}} \left( e^{-\frac{1}{4a} \left( p_1 + 2ah_1 i \right)^2 - ah_1^2 } e^{-\frac{1}{4b} \left( p_2 + 2bh_2 i \right)^2 - bh_2^2 } + e^{-\frac{1}{4a} \left( p_1 - 2ah_1 i \right)^2 - ah_1^2 } e^{-\frac{1}{4b} \left( p_2 - 2bh_2 i \right)^2 - bh_2^2 } \right),
\label{key-7}
\end{eqnarray}
and computing the probability density of $p_1$:
\begin{eqnarray}
f_{p_1} \left(p_1\right) = \pi G \sqrt{\frac{2 \pi}{a}} e^{- \frac{p_1^2}{2a}} \left[ 1 + e^{- 2bh_2^2} \cos \left( 2h_1 p_1 \right) \right].
\label{f_p1_asym}
\end{eqnarray}
As before, \eqref{f_p1_asym} yields an expression for the upper (lower) envelope by substituting $ \cos \left( 2h_1 p_1 \right) = +1 \left( -1 \right) $
\begin{align}
env_{asymmetric}^\pm = \pi G \sqrt{\frac{2 \pi}{a}} e^{- \frac{p_1^2}{2a}} \left[ 1 \pm e^{- 2bh_2^2} \right].
\end{align}
As in the previous section, these are approximations, valid for $ ah_1^2 \gg 1 $; however, we need not assume anything about $b, h_2$. Again, we obtain a constant ratio
\begin{equation}
r_{asymmetric} = \frac{1 + e^{-2bh_2^2} }{1 -e^{-2bh_2^2} } = \coth \left( bh_2^2 \right),
\label{key-8}
\end{equation}
and constant visibility
\begin{equation}
V_{asymmetric} = e^{-2bh_2^2}.
\label{V_asym}
\end{equation}
Note that because of the asymmetry, the one-particle visibility for particle $1$ (which is the one defined above) differs from the one-particle visibility for particle $2$.
Comparing \eqref{V_asym} and \eqref{V_theta}, we wish to find a relation between the parameters determining the one-particle interference visibility. In the ``asymmetric'' wavefunction, it is determined by $b,h_2$, while in the theta-wavefunction it is determined by $a, h_1, \sin \left( 2 \theta \right)$. Taking the two expressions to be equal, we obtain
\begin{equation}
\frac{ e^{-2ah_1^2} + \sin \left( 2\theta \right)}{ 1 + \sin \left( 2\theta \right) e^{-2ah_1^2} } = e^{-2bh_2^2}.
\label{V_theta_eq_V_asym}
\end{equation}
Solving for $\sin \left(2\theta\right)$ yields
\begin{equation}
\sin \left(2\theta\right) = \frac{e^{-2bh_2^2} - e^{-2ah_1^2}}{ 1 - e^{-2ah_1^2} e^{-2bh_2^2} } = \frac{V_{asymmetric} - e^{-2ah_1^2} }{ 1 - e^{-2ah_1^2} V_{asymmetric} }.
\label{theta_vs_b_l}
\end{equation}
This also allows a simple way for Alice to generate a \emph{purification} of her state, using only the parameters of her system and the interference visibility she measures. To put it more precisely, the state of the form \eqref{psi_theta} where $\theta$ is given by \eqref{theta_vs_b_l}, yields the same single-particle Wigner function for Alice as the state \eqref{asym} considered throughout this section (a proof appears in the Appendix \ref{Wigner_relation}). Thus, \eqref{theta_vs_b_l} allows Alice to utilize any of the results obtained in the previous section, e.g. by computing the predictability.

We also note that Alice and Bob may switch roles and achieve similar results; this fact is manifested in \eqref{theta_vs_b_l} by noting the expression is invariant under permutation of the two particles.

%%%%%%%%%%%%%%%%%%%%%%%%%%%%%%%%%%%%%%%%%%
\section{Conclusions}
In this work we derived a complementarity relation analogous to the one presented in \cite{jaeger1993complementarity}, for a family of CV systems described by EPR-like states.
This complementarity relation is sensitive to the parameters of the double-double-slit experiment, namely, the width of each double slit and the separation between the slits. In addition, we have demonstrated that in CV systems, the ``analog'' visibility measures defined using interferometric envelopes, are closely related to the ones defined in previous works \cite{jaeger1993complementarity,greenberger1988simultaneous,jaeger1995two,englert1996fringe} with binary-outcome quantum probabilities.

Moreover, \eqref{theta_vs_b_l} allows application of this complementarity relation to any experiment where a CV is subjected to a binary-outcome measurement: one should have it evolve in a manner analogous to propagation of the double-slit wavefunction to the far-field; afterwards, an interference pattern should be extracted, either an actual one or some analogue; and finally, computation of the envelope-based visibility allow finding the ``$\theta$'' and consequentially the predictability, thus inferring the strength of entanglement.

Additionally, common descriptions of the double-slit experiment usually consider two variants: one where a detector is placed next to one of the slits, and another variant where no detector is placed. However, since the process of measurement may be described by entangling the system and the measuring pointer, a natural way to describe \emph{weak} measurements \cite{AAV88,TC13,Dressel14,vaidman2017weak} or intermediate strength measurements \cite{piacentini2018investigating,dziewior2019universality} in a double-slit setting is by considering weak entanglement of the system and pointer. In this context, by varying the parameter $\theta$ which controls the amount of entanglement between the two particles in the initial state $\psi_{\theta} (x_1,x_2)$, this work may be also viewed as a model for ``weak'' variants of the double-slit experiment with continuous variables.

Finally, this paper strengthens and elucidates the tight relation between local and nonlocal correlations that we previously explored e.g. in \cite{carmi2019relativistic,carmi2018significance,carmi2019bounds}.

\appendix
\section{Complementary quantum measures and relations}\label{measures}

Quantum particles are able to explore simultaneously multiple paths in a quantum
coherent superposition. In general, the $n$ alternative paths could
be explicitly described with the use of quantum histories $\hat{Q}_{1}$,
$\hat{Q}_{2}$, $\ldots$, $\hat{Q}_{n}$, which are defined in history Hilbert space as tensor products
of projection operators at multiple times, and the corresponding quantum
probability amplitudes $\psi_{1}$, $\psi_{2}$, $\ldots$, $\psi_{n}$
propagating along each path can be experimentally measured in the
form of complex-valued sequential weak values \cite{Georgiev2018}.
Without loss of generality, after normalization it may be assumed
that all given paths form a complete set of quantum histories $\sum_{i}|\psi_{i}|^{2}=1$.
In the special case of $n$-slit interference setups, the sequential
weak values reduce to ordinary weak values \cite{AAV88,TC13} with
a single intermediate time at the slits, between the initial emission
from the particle source and the final detection at a sensitive screen.
The quantum \emph{coherence} $\mathcal{C}$ exhibited by each quantum
particle in multi-slit setups could be manifested in \emph{visibility} $\mathcal{V}$
of interference fringes and is constrained by complementary relations to path \emph{predictability}
$\mathcal{P}$ or path \emph{distinguishability} $\mathcal{D}$,
the latter demanding quantum entanglement with external path measuring
devises.
\begin{defn}
	Path \emph{distinguishability} $\mathcal{D}$ is defined with the
	use of external entanglement with measuring devices inserted on the
	paths. For the case of $n$ paths, the distinguishability is \cite{Zhang2001,Qiu2002,Bera2015,Qureshi2017}
	\begin{equation}
	\mathcal{D}=\sqrt{1-\left(\frac{1}{n-1}\sum_{i\neq j}|\psi_{i}||\psi_{j}||\langle d_{i}|d_{j}\rangle|\right)^{2}}\label{eq:D}
	\end{equation}
	where different device states $|d_{i}\rangle$, $|d_{j}\rangle$ are
	normalized but not necessarily orthogonal \cite{Siddiqui2016}. For
	orthogonal states $\langle d_{i}|d_{j}\rangle=0$, the path-distinguishability
	is maximal, $\mathcal{D}=1$.
	
	For two equally likely paths, the path distinguishability reduces
	to
	\begin{equation}
	\mathcal{D}=\sqrt{1-|\langle d_{1}|d_{2}\rangle|^{2}}
	\end{equation}
\end{defn}

\begin{defn}
	Path \emph{predictability} $\mathcal{P}$ is defined without the need
	of external entanglement. The idea is that if two paths have different
	probabilities, one could predict which of the two paths has been taken
	with a success which exceeds a fifty percent random guess. For $n$
	paths, the predictability is \cite{Roy2019}
	\begin{equation}
	\mathcal{P}=\sqrt{1-\left(\frac{1}{n-1}\sum_{i\neq j}|\psi_{i}||\psi_{j}|\right)^{2}}\label{eq:P}
	\end{equation}
	
	For two paths, the path-predictability becomes \cite{greenberger1988simultaneous}
	\begin{equation}
	\mathcal{P}=\left||\psi_{1}|^{2}-|\psi_{2}|^{2}\right|\label{eq:GY}
	\end{equation}
\end{defn}
\begin{proof}
	Indeed, squaring both sides of (\ref{eq:P}) and using $\sum_{i}|\psi_{i}|^{2}=1$,
	we obtain
	\begin{eqnarray*}
		\mathcal{P}^{2} & = & 1^{2}-\left(2|\psi_{1}||\psi_{2}|\right)^{2}=\left(|\psi_{1}|^{2}+|\psi_{2}|^{2}\right)^{2}-4|\psi_{1}|^{2}|\psi_{2}|^{2}\\
		& = & |\psi_{1}|^{4}+2|\psi_{1}|^{2}|\psi_{2}|^{2}+|\psi_{2}|^{4}-4|\psi_{1}|^{2}|\psi_{2}|^{2}\\
		& = & \left(|\psi_{1}|^{2}-|\psi_{2}|^{2}\right)^{2}
	\end{eqnarray*}
	Taking the square root on both sides gives (\ref{eq:GY}).
\end{proof}

\begin{defn}
	Path \emph{coherence} $\mathcal{C}$ quantifies the quantum interference
	between different paths \cite{Bera2015}
	\begin{equation}
	\mathcal{C}=\frac{1}{n-1}\sum_{i\neq j}|\psi_{i}||\psi_{j}||\langle d_{i}|d_{j}\rangle|\label{eq:C}
	\end{equation}
	In the absence of any path detectors, $\langle d_{i}|d_{j}\rangle=1$,
	the coherence reduces to
	\begin{equation}
	\mathcal{C}=\frac{1}{n-1}\sum_{i\neq j}|\psi_{i}||\psi_{j}|
	\end{equation}
\end{defn}

\begin{defn}
	Fringe \emph{visibility} $\mathcal{V}$ quantifies the contrast in
	observed interferometric distribution patterns and is given by
	\begin{equation}
	\mathcal{V}=\frac{I_{\max}-I_{\min}}{I_{\max}+I_{\min}}\label{eq:V}
	\end{equation}
	For two paths, the fringe visibility reduces to coherence
	\begin{equation}
	\mathcal{V}=\frac{2|\psi_{1}||\psi_{2}|}{|\psi_{1}|^{2}+|\psi_{2}|^{2}}=2|\psi_{1}||\psi_{2}|
	\end{equation}
\end{defn}
\begin{proof}
	Consider using (\ref{eq:V}) together with the Born rule for $I_{\max}$
	and $I_{\min}$ to get
	\begin{eqnarray*}
		\mathcal{V} & = & \frac{\left(|\psi_{1}|+|\psi_{2}|\right)^{2}-\left(|\psi_{1}|-|\psi_{2}|\right)^{2}}{\left(|\psi_{1}|+|\psi_{2}|\right)^{2}+\left(|\psi_{1}|-|\psi_{2}|\right)^{2}}\\
		& = & \frac{|\psi_{1}|^{2}+|\psi_{2}|^{2}+2|\psi_{1}||\psi_{2}|-\left(|\psi_{1}|^{2}+|\psi_{2}|^{2}-2|\psi_{1}||\psi_{2}|\right)}{|\psi_{1}|^{2}+|\psi_{2}|^{2}+2|\psi_{1}||\psi_{2}|+|\psi_{1}|^{2}+|\psi_{2}|^{2}-2|\psi_{1}||\psi_{2}|}\\
		& = & \frac{4|\psi_{1}||\psi_{2}|}{2|\psi_{1}|^{2}+2|\psi_{2}|^{2}}=2|\psi_{1}||\psi_{2}|
	\end{eqnarray*}
\end{proof}

Even though for double-slit setups the fringe visibility reduces to
coherence, for multiple slits fringe visibility is not a good measure
of coherence. In fact, fringe visibility may increase with increasing
decoherence as explicitly shown in a specific case of four-path interference
by Qureshi \cite{Qureshi2019}.

The most general equality for multi-path interference of a quantum
particle in the presence of path detectors is between path distinguishability
and path coherence
\begin{equation}
\mathcal{D}^{2}+\mathcal{C}^{2}=1
\end{equation}

\begin{proof}
	Sum the squares of (\ref{eq:D}) and (\ref{eq:C}) to verify that
	the result is a unit.
\end{proof}
As a special case, where there are no external path measuring devices,
$\langle d_{i}|d_{j}\rangle=1$, we obtain path relationship between
path predictability and path coherence
\begin{equation}
\mathcal{P}^{2}+\mathcal{C}^{2}=1
\end{equation}
And if we further limit the setup only to two-path interference, the
coherence can be replaced with fringe visibility
\begin{equation}
\mathcal{D}^{2}+\mathcal{V}^{2}=1
\end{equation}
\begin{equation}
\mathcal{P}^{2}+\mathcal{V}^{2}=1
\end{equation}

Recently, Qureshi proposed to assess $n$ path coherence using a method
of blocking most of the paths in order to measure interference visibilities
of path pairs \cite{Qureshi2019b}. It should be noted that the moduli
of quantum probability amplitudes $|\psi_{i}|$ for $n$ paths could
be determined with only $n$ measurements of individual path probabilities,
$|\psi_{i}|=\sqrt{P_{i}}$. However, to determine orthogonality relations
between external path detectors $\left|\langle d_{i}|d_{j}\rangle\right|$one
would need $n^{2}-n$ pairs of measurements. In essence, Qureshi's
proposal is to combine both types of measurements and assess $n^{2}-n$
pairs of visibilities $|\psi_{i}||\psi_{j}||\langle d_{i}|d_{j}\rangle|$,
which will then be summed according to formula (\ref{eq:C}).

\section{Computations and Wigner functions}
\unskip
\subsection{The Wigner function for a partially-entangled double-slit experiment}\label{wigner_partial}
Let us write down the Wigner function of the state \eqref{psi_theta}:
\begin{align}
W_{\theta} \left( x_1, x_2, p_1, p_2 \right) = & B_{\theta} \cos^2 \theta \left( g_1^- g_2^- +  g_1^+ g_2^+ + 2 g_1^0 g_2^0 \cos \left[ 2h \left( p_1 + p_2 \right) \right] \right) + \nonumber\\
& + B_{\theta} \sin \left( 2 \theta \right) \left[ \cos \left( 2h p_1 \right) g_1^0 \left( g_2^- + g_2^+ \right) + \cos \left( 2h p_2 \right) g_2^0 \left(
g_1^- + g_1^+ \right) \right] \nonumber\\ &
+ B_{\theta} \sin^2 \theta \left( g_1^- g_2^+ + g_1^+ g_2^- + 2 g_1^0 g_2^0 \cos \left[ 2h \left( p_1 - p_2 \right) \right] \right)
\end{align}
where $ g_i^\pm \defeq e^{- 2a \left( x_i \pm h \right)^2} e^{- \frac{p_i^2}{ 2a } } $, $ g_i^0 \defeq e^{- 2a x_i^2} e^{- \frac{p_i^2}{ 2a } } $ and $ B_{\theta} \defeq \frac{A_{\theta}^2 }{2 \pi a} $.
Now we shall compute the ``partial'' Wigner function of $x_1, p_1$
\begin{align}\label{W_1_theta}
& W_{1, \theta} \left( x_1, p_1 \right) \defeq \int dx_2 \int dp_2 \, W_{\theta} \left( x_1, x_2, p_1, p_2 \right) = \nonumber\\
& = \pi B_{\theta} \left[ 1 + e^{-2ah^2} \sin \left( 2\theta \right) \right] \left( g_1^- + g_1^+ \right) + 2 \pi B_{\theta} \left[ e^{-2ah^2} + \sin \left( 2\theta \right) \right] g_1^0 \cos \left( 2h p_1 \right)
\end{align}
Note that, for $ \theta = \pi / 4 $ we have
\begin{eqnarray}
&& W_{1, \theta = \pi / 4} \left( x_1, p_1 \right) =
\frac{1}{2 \pi \left( 1 + e^{-2ah^2} \right) } \left[ g_1^- + g_1^+ + 2 g_1^0 \cos \left( 2hp_1 \right) \right]
\end{eqnarray}
which is equal to the Wigner function for a single-particle double-slit system.
The momentum probability density is computed by
\begin{align}
f_{p_1, \theta} \left( p_1 \right) & = \int dx_1 \, W_{1, \theta} \left( x_1, p_1 \right)
\end{align}
and the momentum \emph{joint} probability distribution is
\begin{align}
& f_{p_1 p_2, \theta} \left( p_1, p_2 \right) \defeq \int dx_1 \int dx_2 \, W_\theta \left( x_1, x_2, p_1, p_2 \right) = \\
& = B_\theta \frac{\pi}{a} e^{- \frac{p_1^2 + p_2^2 }{ 2a} } \left( 1+ \cos^2 \theta \cos \left[ 2h \left( p_1 +p_2 \right) \right] +\sin \left( 2 \theta \right) \left[ \cos \left( 2h p_1 \right) + \cos \left( 2h p_2 \right) \right] + \sin^2 \theta  \cos \left[ 2h \left( p_1 -p_2 \right) \right] \right) = \nonumber\\
& = B_\theta \frac{\pi}{a} e^{- \frac{p_1^2 + p_2^2 }{ 2a} } \left( 1+ \cos \left( 2h p_1 \right) \cos \left( 2h p_2 \right) -\cos \left( 2\theta \right) \sin \left( 2h p_1 \right) \sin \left( 2h p_2 \right) + \sin \left( 2 \theta \right) \left[ \cos \left( 2h p_1 \right) + \cos \left( 2h p_2 \right) \right] \right) = \nonumber\\
& = B_\theta \frac{\pi}{a} e^{- \frac{p_1^2 + p_2^2 }{ 2a} } \left[ 1+ \cos^2 \theta \cos \left( 2h p_+ \right) +2 \sin \left( 2 \theta \right) \cos \left( h p_+ \right)  \cos \left( h p_- \right) + \sin^2 \theta  \cos \left( 2h p_- \right) \right] .
\end{align}
The ``corrected'' momentum joint probability is defined by
\begin{align}
\tilde{F}_\theta \left( p_1, p_2 \right) \defeq & f_{p_1 p_2, \theta} \left( p_1, p_2 \right) - f_{p_1, \theta} \left( p_1 \right) f_{p_2, \theta} \left( p_2 \right) + \frac{B_\theta}{B_{\theta=0}} f_{p_1, \theta=0} \left( p_1 \right) f_{p_2, \theta=0} \left( p_2 \right) = \nonumber\\
= & + \frac{e^{-\frac{p_1^2 + p_2^2}{2a} } }{ C_\theta } \left[ e^{4 a h^2} + \cos \left(2 h p_1 \right) \cos \left(2 h p_2 \right) + \cos^2 \left( 2\theta \right) \right] + \nonumber\\
& + \frac{e^{-\frac{p_1^2 + p_2^2}{2a} } }{ C_\theta } \left[ \sin^2 \left( 2\theta \right) e^{2 a h^2} \left[ \cos \left(2 h p_1 \right) + \cos \left(2 h p_2 \right) \right] -\cos \left( 2\theta \right) \sin \left(2 h p_1\right) \sin \left(2 h p_2\right) \right] + \nonumber\\
& +\frac{e^{-\frac{p_1^2 + p_2^2}{2a} } }{ C_\theta } \left[ \cos^2 \left( 2\theta \right) e^{4 a h^2} \cos \left(2 h p_1 \right) \cos \left(2 h p_2 \right) - \cos \left( 2\theta \right) e^{4 a h^2} \sin \left(2 h p_1\right) \sin \left(2 h p_2\right) \right] +\nonumber\\
& -\frac{e^{-\frac{p_1^2 + p_2^2}{2a} } }{ C_\theta } \sin \left( 4\theta \right) e^{2 a h^2} \sin \left(2 h p_1\right) \sin \left(2 h p_2\right) = \nonumber\\
= & + \frac{e^{-\frac{p_+^2 + p_-^2}{4a} } }{ C_\theta } \left( e^{4 a h^2} + \cos^2 \left( 2\theta \right) +\frac{1}{2} \left[ 1+ \cos^2 \left( 2\theta \right) e^{4ah^2} \right] \left[ \cos \left( 2h p_+ \right) + \cos \left( 2h p_- \right) \right] \right) + \nonumber\\ & +2 \frac{e^{-\frac{p_+^2 + p_-^2}{4a} } }{ C_\theta } \sin^2 \left( 2 \theta \right) e^{2ah^2} \cos \left( h p_+ \right) \cos \left( h p_- \right) + \nonumber\\ & - \frac{1}{2} \frac{e^{-\frac{p_+^2 + p_-^2}{4a} } }{ C_\theta } \cos \left( 2 \theta \right) \left[ 1+ e^{4ah^2} + 2 \sin \left( 2 \theta \right) \right] \left[ \cos \left( 2h p_- \right) - \cos \left( 2h p_+ \right) \right] = \nonumber\\
= & + \frac{e^{-\frac{p_+^2 + p_-^2}{4a} } }{ C_\theta } \left[ e^{4 a h^2} + \cos^2 \left( 2\theta \right) +2 \sin^2 \left( 2 \theta \right) e^{2ah^2} \cos \left( h p_+ \right) \cos \left( h p_- \right) \right] + \nonumber\\
& + \frac{1}{2} \frac{e^{-\frac{p_+^2 + p_-^2}{4a} } }{ C_\theta } \left( 1+ \cos^2 \left( 2\theta \right) e^{4ah^2} + \cos \left( 2 \theta \right) \left[ 1+ e^{4ah^2} + 2 \sin \left( 2 \theta \right) \right] \right) \cos \left( 2h p_+ \right) + \nonumber\\
& + \frac{1}{2} \frac{e^{-\frac{p_+^2 + p_-^2}{4a} } }{ C_\theta } \left( 1+ \cos^2 \left( 2\theta \right) e^{4ah^2} - \cos \left( 2 \theta \right) \left[ 1+ e^{4ah^2} + 2 \sin \left( 2 \theta \right) \right] \right) \cos \left( 2h p_- \right) .
\label{key}
\end{align}
Substituting $p_- = 0$, we obtain
\begin{align}
\tilde{F}_\theta \left( p_+, p_- = 0 \right) & = \frac{e^{-\frac{p_+^2}{4a} } }{ C_\theta } \left[ e^{4 a h^2} + \cos^2 \left( 2\theta \right) +2 \sin^2 \left( 2 \theta \right) e^{2ah^2} \cos \left( h p_+ \right) \right] + \nonumber\\
& + \frac{1}{2} \frac{e^{-\frac{p_+^2}{4a} } }{ C_\theta } \left( 1+ \cos^2 \left( 2\theta \right) e^{4ah^2} + \cos \left( 2 \theta \right) \left[ 1+ e^{4ah^2} + 2 \sin \left( 2 \theta \right) \right] \right) \cos \left( 2h p_+ \right) + \nonumber\\
& + \frac{1}{2} \frac{e^{-\frac{p_+^2}{4a} } }{ C_\theta } \left( 1+ \cos^2 \left( 2\theta \right) e^{4ah^2} - \cos \left( 2 \theta \right) \left[ 1+ e^{4ah^2} + 2 \sin \left( 2 \theta \right) \right] \right) .
\end{align}

\subsection{The Wigner function for an asymmetric double-slit experiment}\label{Wigner_asym}
The Wigner function corresponding to \eqref{asym} is:
\begin{align}\label{Wigner_maximally_entangled_asym}
W \left(x_1, x_2, p_1, p_2 \right) =
G \left( \gamma_1^- \gamma_2^- + \gamma_1^+ \gamma_2^+ \right) +
2G \gamma_1^0 \gamma_2^0 \cos \left( 2h_1 p_1 + 2h_2 p_2 \right)
\end{align}
where $ \gamma_1^\pm \defeq e^{ -2a \left( x_1 \pm h_1 \right)^2} e^{- \frac{p_1^2}{ 2a } } $, $ \gamma_2^\pm \defeq e^{ -2b \left( x_2 \pm h_2 \right)^2} e^{- \frac{p_2^2}{ 2b } } $, $ \gamma_1^0 \defeq e^{ -2a x_1^2} e^{- \frac{p_1^2}{ 2a } } $, $ \gamma_2^0 \defeq e^{ -2b x_2^2} e^{- \frac{p_2^2}{ 2b } } $ and $ G \defeq \frac{M^2}{2\pi \sqrt{ab}} $.
The ``partial'' Wigner function is:
\begin{align}\label{W1_asym}
W_1 \left( x_1, p_1 \right) \defeq \int dx_2 \int dp_2 \, W \left( x_1, x_2, p_1, p_2 \right) =
\pi G \left[ \gamma_1^- + \gamma_1^+ + 2 e^{-2b h_2^2} \gamma_1^0 \cos \left( 2 h_1 p_1 \right) \right] .
\end{align}

\subsection{Purification of Alice's subsystem\label{Wigner_relation}}

Suppose Alice is entangled with Bob in an asymmetric manner, and has no knowledge of anything outside her system; however, she knows the parameters $a, h_1$. She performs interference experiments, obtains an interference pattern and uses it to compute $V_{asymmetric}$. Here we demonstrate she can always \emph{purify} her state - i.e., write down a pure ``Alice-Bob'' state of the form \eqref{psi_theta} yielding the same one-particle Wigner function for Alice.

Our claim here, is that the state \eqref{psi_theta} where $\theta$ is a solution of \eqref{theta_vs_b_l} satisfies the above requirement. Indeed, substituting \eqref{V_theta_eq_V_asym} in \eqref{W1_asym} yields:
\begin{align}
& W_{1,asym} \left( x_1, p_1 \right) = \pi G \left[ g_1^- + g_1^+ + 2 g_1^0 \frac{ e^{-2ah_1^2} + \sin \left( 2\theta \right)}{ 1 + \sin \left( 2\theta \right) e^{-2ah_1^2} } \cos \left( 2 h_1 p_1 \right) \right] = \nonumber\\
& = \frac{\pi G}{1 + \sin \left( 2\theta \right) e^{-2ah_1^2}}
\left( \left[ 1 + e^{-2ah_1^2} \sin \left( 2\theta \right) \right] \left( g_1^- + g_1^+ \right) +2 \left[ e^{-2ah_1^2} + \sin \left( 2\theta \right) \right] g_1^0 \cos \left( 2 h_1 p_1 \right) \right)
\label{key-0}
\end{align}
which is identical to \eqref{W_1_theta} up to normalization, implying it should be completely identical, thus concluding the proof.

\bibliographystyle{unsrt}  
\bibliography{ArticleReview}  %%% Remove comment to use the external .bib file (using bibtex).


\end{document}